\documentclass[pre,aps,amsfonts,amssymb,showpacs,preprint]{revtex4}

\usepackage{graphicx,bm}

\def\u{{\bm u}}

\def\f{{\bm f}}

\def\norm#1{\left|\mkern-2mu\left|#1\right|\mkern-2mu\right|}

\begin{document}

\title{Number of degrees of freedom of two-dimensional turbulence}

\author{Chuong V. Tran and Luke Blackbourn}

\affiliation{School of Mathematics and Statistics, University of St Andrews,
St Andrews KY16 9SS, United Kingdom}

\date{\today}

\begin{abstract}

We derive upper bounds for the number of degrees of freedom of 
two-dimensional Navier--Stokes turbulence freely decaying from a smooth
initial vorticity field $\omega(x,y,0)=\omega_0$. This number, denoted 
by $N$, is defined as the minimum dimension such that for $n\ge N$, 
arbitrary $n$-dimensional balls in phase space centred on the solution 
trajectory $\omega(x,y,t)$, for $t>0$, contract under the dynamics of 
the system linearized about $\omega(x,y,t)$. In other words, $N$ is the 
minimum number of greatest Lyapunov exponents whose sum becomes negative. 
It is found that $N\le C_1R_e$ when the phase space is endowed with the 
energy norm, and $N\le C_2R_e(1+\ln R_e)^{1/3}$ when the phase space is 
endowed with the enstrophy norm. Here $C_1$ and $C_2$ are constant and 
$R_e$ is the Reynolds number defined in terms of $\omega_0$, the system 
length scale, and the viscosity $\nu$. The linear (or nearly linear) 
dependence of $N$ on $R_e$ is consistent with the estimate for the number
of active modes deduced from a recent mathematical bound for the viscous 
dissipation wave number. This result is in a sharp contrast to the forced 
case, for which well-known estimates for the Hausdorff dimension $D_H$ 
of the global attractor scale highly superlinearly with $\nu^{-1}$. We 
argue that the ``extra'' dependence of $D_H$ on $\nu^{-1}$ is not an 
intrinsic property of the turbulent dynamics. Rather, it is a ``removable 
artifact,'' brought about by the use of a time-independent forcing 
as a model for energy and enstrophy injection that drives the turbulence.

\end{abstract}


\maketitle
\widetext
\section{INTRODUCTION}

Chaotic dynamics are characterized by the stretching and folding 
of volume elements in phase space (solution space). In the presence 
of dissipation, these can be accompanied by volume contraction. For a
finite-dimensional system, volume elements can eventually collapse onto 
complex sets of zero volume having fractal structures, whose generalized
dimensions, such as the box-counting and Hausdorff dimensions, are 
significantly lower than the phase space dimension. For 
infinite-dimensional systems, volume contraction can occur for 
finite-dimensional volume elements. Furthermore, given a sufficiently 
large positive integer $N$ (depending on physical parameters and initial
conditions), this contraction can occur for arbitrarily oriented 
$n$-dimensional volume elements following a trajectory --- solution 
``curve'' in function phase space --- provided that $n\ge N$. This is the 
case if the sum of the largest $N$ Lyapunov exponents at each point of 
the trajectory $\lambda_1\ge\lambda_2\ge\cdots\ge\lambda_N$, which can 
possibly be different for different trajectories, is negative. The smallest 
$N$ (which will be denoted by $N$ still) satisfying this condition thus 
defines the minimum dimension in phase space for which all $n$-dimensional 
($n\ge N$) volume elements along a given trajectory contract during the 
course of evolution. This volume contraction means that the chaotic nature 
of the local dynamics can be ``captured'' and ``contained'' within a linear 
subspace having dimension not higher than $N$. (This subspace may 
continuously change along the trajectory, though its dimension does not 
exceed $N$.) For this reason, $N$ can be thought of as an effective 
dimension of the dynamical system in question, 
in the sense that its local dynamics can be adequately described by an 
$N$-dimensional model. When an attractor (or a global attractor) exists 
and $N$ is common to 
every trajectory having initial data containing the attractor, its 
box-counting and Hausdorff dimensions are both bounded from above by $N$ 
\cite{Hunt96}, which is conveniently defined as the number of degrees 
of freedom. More precisely, these are bounded from above by the 
Lyapunov dimension $D_L$, which satisfies $N-1\le D_L<N$ and is defined 
by \cite{Kaplan79,Farmer82}
\begin{eqnarray}
\label{dlyapunov}
D_L &=& N - 1 + \frac{1}{|\lambda_N|}\sum_{i=1}^{N-1}\lambda_i.
\end{eqnarray}

In this study we determine upper bounds for $N$ for two-dimensional 
Navier--Stokes turbulence freely decaying from a smooth initial vorticity
field $\omega_0$ in a doubly periodic domain of length scale $L$. Note that
the global attractor for this case is trivial and has zero dimension. 
However, the present problem is nontrivial because it is concerned with 
transient dynamics, most importantly during the stage of fully developed 
turbulence. The bounds obtained are expressible in terms of physical 
parameters and 
found to scale linearly or almost linearly (depending on the chosen norms 
for the phase space) with the Reynolds number $R_e$, which is defined in 
terms of $\omega_0$, $L$, and the viscosity $\nu$. On the one hand, such 
scaling behaviors are in accord with heuristic arguments based on physical 
and mathematical estimates of the viscous dissipation wave number. On the 
other hand, these are in a sharp contrast to the forced case, for which 
well-known upper bounds for the Hausdorff dimension $D_H$ of the global 
attractor have superlinear dependence on $\nu^{-1}$ 
\cite{Babin83,Constantin85,Constantin88}. We discuss 
this discrepancy and argue that the superlinear dependence of $D_H$ on 
$\nu^{-1}$ is not an intrinsic property of the turbulent dynamics. Rather,
it appears to be a ``removable artifact,'' brought about by the particular 
form of the forcing term used as a model for energy and enstrophy injection 
that drives the turbulence. Indeed, the ``extra'' dependence of $D_H$ on 
$\nu^{-1}$ would be removed if the energy and enstrophy injection could
be made viscosity independent (more precisely if the injection could be 
bounded independently of viscosity), provided that this forcing model does
not jeopardize the existence of the global attractor. 
 
\section{Preliminaries}

In this section we briefly recall a recently derived upper bound 
\cite{T05} for the enstrophy dissipation wave number $k_d$. We deduce 
from this result an estimate for the number of active modes, by counting 
all modes having length scales larger than the dissipation length scale 
corresponding to $k_d$. A brief functional setting of the 
two-dimensional Navier--Stokes system in the vorticity and stream 
function formulation is described, and the problem of phase space volume 
evolution is formulated. We avoid technical detail and use informal
language.

\subsection{Number of active modes}

The two-dimensional Navier--Stokes system written in terms of the 
stream function $\psi$ and vorticity $\omega=\Delta\psi$ is
\begin{eqnarray}
\label{vorticity} 
\omega_t + J(\psi,\omega) = \nu\Delta\omega,
\end{eqnarray}
where $J(\psi,\omega)=\psi_x\omega_y - \psi_y\omega_x$ and $\nu$ is the
viscosity. We consider Eq.\ (\ref{vorticity}) in a doubly periodic domain
of size $2\pi L$. The initial vorticity field $\omega_0$ is assumed to be 
smooth and have zero average. Equation (\ref{vorticity}) preserves the
zero-mean property. This, together with periodicity, allows $\omega$
(and $\psi$) to be expressible as a Fourier series in terms of 
$\sin L^{-1}(\ell x+my)$ and $\cos L^{-1}(\ell x+my)$, where $\ell$ and 
$m$ are integers not simultaneously zero. In other words, the 
infinite-dimensional space (solution function space) can be spanned by 
the infinite basis $\{\sin L^{-1}(\ell x+my),\cos L^{-1}(\ell x+my)\}$.
Here $(\ell,m)$ can be identified with a lattice of unit spacing on the 
upper half plane, with points on either half of the horizontal axis, 
including the origin, removed.

The advection term $J(\psi,\omega)$ has many conservation laws. 
In particular, the total kinetic energy $\norm{\nabla\psi}^2/2=
\langle|\nabla\psi|^2\rangle/2=\int|\nabla\psi|^2\,dxdy/2$, the total 
enstrophy $\norm{\omega}^2/2$, and the peak vorticity 
$\norm{\omega}_\infty$ are conserved. These are the most important 
conserved quantities and play prominent roles in the theory of 
turbulence. Under viscous effects, all these quantities decay, though 
in general at different rates. The enstrophy decays most rapidly, while 
the kinetic energy and the peak vorticity are far better conserved, 
with the latter probably best conserved \cite{DTS07}. 

For relatively small $\nu$, the free decay of a general smooth vorticity 
field presumably becomes turbulent, featuring a wide range of dynamically 
interacting scales that extend to the viscous dissipation range. This 
range is characterized by the dissipation wave number $k_\nu$, which,
according to the phenomenological theory of turbulence \cite{Batchelor69}, 
is given by $k_\nu = \chi^{1/6}/\nu^{1/2}$. Here 
$\chi=\nu\norm{\nabla\omega}^2/(4\pi^2L^2)$ denotes the enstrophy 
dissipation rate per unit area. Recently, Tran \cite{T05} derived the 
upper bound
\begin{eqnarray}
\label{palin}
\nu\norm{\nabla\omega}^2 \le \norm{\omega}_\infty\norm{\omega}^2,
\end{eqnarray}
for the dissipation rate $\nu\norm{\nabla\omega}^2$ at its peak. Since
both vorticity norms on the right-hand side of Eq.\ (\ref{palin}) decay 
we have the bound
\begin{eqnarray}
\label{palins}
\norm{\nabla\omega} \le \frac{\norm{\omega_0}_\infty^{1/2}
\norm{\omega_0}}{\nu^{1/2}},
\end{eqnarray}
which is valid uniformly in time, and the bound
\begin{eqnarray}
\label{kd}
k_d = \frac{\norm{\nabla\omega}}{\norm{\omega}} \le 
\frac{\norm{\omega_0}_\infty^{1/2}}{\nu^{1/2}},
\end{eqnarray}
which is valid at least up to (and probably beyond) the time of peak 
enstrophy dissipation. The bound for the newly defined enstrophy 
dissipation wave number $k_d$ compares favorably to $k_\nu$ as it 
could be significantly smaller than $k_\nu$ \cite{T05}. By the very
definition (\ref{kd}), enstrophy dissipation is strongest in the 
vicinity of $k_d$. The wave numbers greater than $k_d$ are effectively 
suppressed by viscous forces and virtually inactive. The number of 
dynamically active modes $N_c$ corresponding to $k\le k_d$ are therefore 
given by
\begin{eqnarray}
\label{N}
N_c \approx \frac{k_d^2}{k_0^2} \le \frac{L^2\norm{\omega_0}_\infty}{\nu},
\end{eqnarray}
where $k_0=1/L$ is the smallest wave number. The quantity
$L\norm{\omega_0}_\infty$ may be identified with the fluid velocity. 
Perhaps, $\norm{\omega_0}$ is a better representative of the fluid 
velocity; nevertheless, when it comes to the definition of the Reynolds 
number $R_e$, we use $L\norm{\omega_0}_\infty$ and $\norm{\omega_0}$ 
interchangeably. With this identification, the term on the right-hand 
side of Eq.\ (\ref{N}) may be defined as the Reynolds number $R_e$. 
Hence, Eq.\ (\ref{N}) can be rewritten in a more compact form 
\begin{eqnarray}
\label{N1}
N_c \le R_e.
\end{eqnarray}

From our experience in numerical simulations of two-dimensional 
turbulence, the estimate (\ref{N1}) is sharp --- in fact, spot on. 
For example, for the standard numerical domain $2\pi\times2\pi$ and an 
initial vorticity maximum $\norm{\omega_0}_\infty\approx4\pi$, the 
simulations of Dritschel, Tran, and Scott \cite{DTS07} using 
$4\pi(8/3)^2/\nu$ grid points adequately resolve the dissipation scales. 
This resolution is obviously consistent with Eq.\ (\ref{N1}), within an
order of magnitude. As will be seen in the next section, the estimate 
(\ref{N1}) for $N_c$ fully agrees with the number of degrees of freedom 
discussed above.

\subsection{Problem formulation}

The problem of phase space volume element contraction (or expansion) 
is intimately related to the stability of solution with respect to 
disturbances. To investigate this problem, we consider the linear 
evolution of a deviation $\phi$ of the stream function $\psi$ 
(corresponding to a deviation $\Delta\phi$ of the vorticity $\omega$) 
governed by the linearised equation 
\begin{eqnarray}
\label{linearised} 
\Delta\phi_t + J(\phi,\omega) + J(\psi,\Delta\phi) = \nu\Delta^2\phi,
\end{eqnarray} 
where $\omega$ (and $\psi$) solves Eq.\ (\ref{vorticity}) with initial 
vorticity $\omega_0$ (and initial stream function $\psi_0$). By taking 
the scalar product ($\langle\cdot\rangle$) of Eq.\ (\ref{linearised}) 
with $\phi$ and $\Delta\phi$ we obtain the respective evolution equations 
for the energy norm $\norm{\nabla\phi}$ and enstrophy norm 
$\norm{\Delta\phi}$,
\begin{eqnarray} 
\norm{\nabla\phi}\frac{d}{dt}\norm{\nabla\phi} 
= \langle\phi J(\psi,\Delta\phi)\rangle - \nu\norm{\Delta\phi}^2
\end{eqnarray} 
and
\begin{eqnarray} 
\norm{\Delta\phi}\frac{d}{dt}\norm{\Delta\phi} 
= -\langle\Delta\phi J(\phi,\omega)\rangle - \nu\norm{\nabla\Delta\phi}^2.
\end{eqnarray} 
The respective exponential growth (or decay) rates $\lambda$ and 
$\Lambda$ for $\norm{\nabla\phi}$ and $\norm{\Delta\phi}$ can be readily 
deduced and are given by
\begin{eqnarray} 
\label{growthrate1} 
\lambda = \frac{d}{dt}\ln\norm{\nabla\phi} =
\frac{1}{\norm{\nabla\phi}^2}\left(\langle\phi 
J(\psi,\Delta\phi)\rangle - \nu\norm{\Delta\phi}^2\right)
\end{eqnarray} 
and
\begin{eqnarray} 
\label{growthrate2} 
\Lambda = \frac{d}{dt}\ln\norm{\Delta\phi} =
\frac{-1}{\norm{\Delta\phi}^2}\left(\langle\Delta\phi 
J(\phi,\omega)\rangle + \nu\norm{\nabla\Delta\phi}^2\right).
\end{eqnarray} 
These rates provide a comprehensive picture of solution stability, 
quantitatively describing how solutions with nearby initial data 
disperse from one another.

Two natural norms for the present problem are the energy and enstrophy 
norms. We will refer to the phase space equipped with the energy 
(enstrophy) norm as the energy (enstrophy) space. In the course of 
evolution, consider a trajectory commencing from a given initial 
condition. At an arbitrary point on the trajectory (i.e., at an arbitrary
instance in time $t>0$), we calculate the greatest growth rate $\lambda$ 
($\Lambda$) and identify the corresponding most unstable ``direction'' 
by considering the problem of maximizing $\lambda$ ($\Lambda$) with 
respect to all admissible $\phi$. We denote by $(\lambda_1,\varphi_1)$ 
[$(\Lambda_1,\vartheta_1)$] the solution of this problem, where for
convenience $\varphi$ ($\vartheta$) has been normalized, i.e., 
$\norm{\nabla\varphi_1}=1$ ($\norm{\Delta\vartheta_1}=1$). 
The second greatest rate $\lambda_2$ ($\Lambda_2$) and the corresponding 
second most unstable direction $\varphi_2$ ($\vartheta_2$) orthogonal 
to $\varphi_1$ ($\vartheta_1$) is obtained by the same maximization 
problem subject to the orthogonality constraint, i.e., 
$\langle\nabla\varphi_1\cdot\nabla\varphi_2\rangle=0$ 
($\langle\Delta\vartheta_1\Delta\vartheta_2\rangle=0$). 
By repeating this procedure $n$ times, we obtain the set 
$\{\varphi_1,\varphi_2,\cdots,\varphi_n\}$ 
($\{\vartheta_1,\vartheta_2,\cdots,\vartheta_n\}$) of mutually orthonormal
functions and the corresponding set of ordered rates 
$\lambda_1\ge\lambda_2\ge\cdots\ge\lambda_n$ 
($\Lambda_1\ge\Lambda_2\ge\cdots\ge\Lambda_n$). These may be defined as
the first $n$ local Lyapunov exponents, and their existence is guaranteed
since the maximization problems are expected to return unique solutions. 
Note that for the conventional Lyapunov exponents, existence can be a 
major issue, even for low-dimensional systems of a few degrees of freedom.

Now, in the linear subspace spanned by 
$\{\varphi_1,\varphi_2,\cdots,\varphi_n\}$
($\{\vartheta_1,\vartheta_2,\cdots,\vartheta_n\}$), consider an 
$n$-dimensional ball $B(\cdot,r)$ of radius $r$ centred at the point 
discussed above. The $n$-dimensional volumes $v$ (in the energy subspace) 
and $V$ (in the enstrophy subspace) of $B(\cdot,r)$ are given by 
$v\propto r^n\norm{\nabla\varphi_1}\norm{\nabla\varphi_2}\cdots
\norm{\nabla\varphi_n}=r^n$ and
$V\propto r^n\norm{\Delta\vartheta_1}\norm{\Delta\vartheta_2}\cdots
\norm{\Delta\vartheta_n}=r^n$, respectively. (See the book of Temam 
\cite{Temam97} for a formal definition of volume based on the related 
concept of exterior product.) The respective equations governing the 
evolution of $v$ and $V$ under the linearised dynamics described by 
Eq.\ (\ref{linearised}) are
\begin{eqnarray}
\label{sum1} 
\frac{d}{dt}\ln v = \sum_{i=1}^{n}\lambda_i = 
\sum_{i=1}^{n}\left(\langle\varphi^i 
J(\psi,\Delta\varphi^i)\rangle - \nu\norm{\Delta\varphi^i}^2\right)
\end{eqnarray} 
and 
\begin{eqnarray}
\label{sum2} 
\frac{d}{dt}\ln V = \sum_{i=1}^{n}\Lambda_i = -\sum_{i=1}^{n}
\left(\langle\Delta\vartheta^i J(\vartheta^i,\omega)\rangle + 
\nu\norm{\nabla\Delta\vartheta^i}^2\right).
\end{eqnarray}  
In deriving Eqs.\ (\ref{sum1}) and (\ref{sum2}), we have used Eqs.\ 
(\ref{growthrate1}) and (\ref{growthrate2}), respectively. The sum 
$\sum_{i=1}^{n}\lambda_i$ ($\sum_{i=1}^{n}\Lambda_i$) represents the 
exponential growth or decay rate of $v$ ($V$). When this sum is negative, 
the volume of the $n$-dimensional ball $B(\cdot,r)$ contracts exponentially. 
Note that by construction, $B(\cdot,r)$ is optimally ``oriented'' to be 
least contracting. This means that if $\sum_{i=1}^{n}\lambda_i$ 
($\sum_{i=1}^{n}\Lambda_i$) is negative, then volume contraction 
becomes universal for all $n$- or higher-dimensional balls locally
centred at the point in question. Furthermore, if this point is 
taken arbitrarily on the trajectory, which is the case in this study, 
then volume contraction becomes universal along the trajectory.

The determination of $N$ then reduces to minimizing $n$ such that the 
sum on the right-hand side of Eqs.\ (\ref{sum1}) and (\ref{sum2}) is
negative. We use the mathematical techniques developed in the 1980s 
by Babin and Vishik \cite{Babin83} and Constantin, Foias, and Temam 
\cite{Constantin85,Constantin88,Constantin87} for estimating the attractor 
dimension of forced two-dimensional Navier--Stokes turbulence. See also 
the paper of Doering and Gibbon \cite{Doering91} for the same treatment 
in the stream function and vorticity setting. As can be seen in the next
section, the derivation of upper bounds for $N$ is equivalent to the
determination of the Hausdorff dimension of the global attractor in 
the forced case. 
The main difference is that although the present formulation is specifically 
designed to handle the decaying case, which has a trivial global 
attractor, its scope of application is broad. In general, the present 
notion of degrees of freedom makes sense for general dissipative dynamical 
systems, provided that bounded solutions exist. There are virtually no 
other technical requirements for the application of the method. In 
particular, no {\it a priori} knowledge of the existence of an attractor 
is required.

\section{RESULTS}

This section presents the calculations described above, leading to upper 
bounds for $N$. The treatment is relatively self-contained. However, the 
reader, who is interested in further detail related to the analytic 
inequalities employed in various stages of the calculations, is referred 
to the cited papers and references therein.

\subsection{Degrees of freedom in energy space}

We begin by deriving an upper bound for $N$ in the energy space. From
Eq.\ (\ref{sum1}) we have
\begin{eqnarray}
\label{SUM1}
\sum_{i=1}^{n}\lambda_i 
&=& 
\sum_{i=1}^{n}\left(\langle\varphi^i J(\psi,\Delta\varphi^i)\rangle 
- \nu\norm{\Delta\varphi^i}^2\right) \nonumber\\ 
&=&
-\sum_{i=1}^{n}\left(\langle\Delta\varphi^i J(\psi,\varphi^i)\rangle 
+ \nu\norm{\Delta\varphi^i}^2\right) \nonumber\\ 
&=&
-\sum_{i=1}^{n}\left(\langle\varphi^i J(\psi_x,\varphi^i_x) +
\varphi^i J(\psi_y,\varphi^i_y)\rangle 
+ \nu\norm{\Delta\varphi^i}^2\right) \nonumber\\
&=&
\sum_{i=1}^{n}\left(\langle\varphi_x^i J(\psi_x,\varphi^i) +
\varphi_y^i J(\psi_y,\varphi^i)\rangle 
- \nu\norm{\Delta\varphi^i}^2\right) \nonumber\\ 
&\le&
\sum_{i=1}^{n}\left(\langle|\nabla\varphi^i|(|\varphi^i_x||\nabla\psi_x|
+ |\varphi^i_y||\nabla\psi_y|)\rangle
- \nu\norm{\Delta\varphi^i}^2\right) \nonumber\\ 
&\le&
\sum_{i=1}^{n}\left(\langle|\nabla\varphi^i|^2
(|\nabla\psi_x|^2+|\nabla\psi_y|^2)^{1/2}\rangle  
- \nu\norm{\Delta\varphi^i}^2\right) \nonumber\\ 
&\le&
\norm{\omega}\norm{\sum_{i=1}^{n}|\nabla\varphi^i|^2} 
- \nu\sum_{i=1}^{n}\norm{\Delta\varphi^i}^2,
\end{eqnarray}
where integration by parts and the Cauchy--Schwarz inequality have been 
used. For further estimates of the terms on the right-hand side of 
Eq.\ (\ref{SUM1}), we employ the following two analytic inequalities
concerning the orthonormal set $\{\varphi^i\}_{i=1}^n$ with respect 
to the energy norm. First, we have the Lieb--Thirring inequality 
\cite{Temam97,Constantin88}
\begin{eqnarray}
\label{ineq1} 
\norm{\sum_{i=1}^{n}|\nabla\varphi^i|^2} \le c_1
\left(\sum_{i=1}^n\norm{\Delta\varphi^i}^2\right)^{1/2},
\end{eqnarray}  
where $c_1$ is a non-dimensional constant independent of the set 
$\{\varphi^i\}_{i=1}^n$. Second, we know that for $n\gg1$, there are
approximately $n$ basis functions (trigonometric functions mentioned 
earlier) within the wave number radius $\sqrt{n}/L$. Their (repeated) 
eigenvalues under $-\Delta$ are $(\ell^2+m^2)/L^2$, where 
$\ell^2+m^2\le n$. These constitute the first $n$ eigenvalues
(in non-decreasing order) of $-\Delta$ and sum up to approximately 
$n^2/L^2$. It follows from the Rayleigh--Ritz principle that
\begin{eqnarray}
\label{ineq2} 
\sum_{i=1}^n\norm{\Delta\varphi^i}^2 \ge \frac{c_2^2}{L^2}n^2,
\end{eqnarray}  
where $c_2$ is another non-dimensional constant independent of the set 
$\{\varphi^i\}_{i=1}^n$. By substituting Eqs.\ (\ref{ineq1}) and
(\ref{ineq2}) into Eq.\ (\ref{SUM1}) we obtain
\begin{eqnarray} 
\sum_{i=1}^{n}\lambda_i &\le&
\left(\sum_{i=1}^{n}\norm{\Delta\varphi^i}^2\right)^{1/2}
\left(c_1\norm{\omega} - \nu\frac{c_2}{L}\,n \right).
\end{eqnarray}  
It follows that $\sum_{i=1}^{n}\lambda_i \le 0$ when 
$n \ge c_1L\norm{\omega}/(c_2\nu)$. Hence we deduce the bound
\begin{eqnarray} 
\label{final1}
N &\le& C_1\frac{L\norm{\omega}}{\nu} \le 
C_1\frac{L\norm{\omega_0}}{\nu} = C_1R_e,
\end{eqnarray} 
where $C_1=c_1/c_2$ and $R_e$ has been redefined by replacing
$L\norm{\omega_0}_\infty$ with $\norm{\omega_0}$. Note that the precise 
result should be that $N$ is no greater than the least integral upper 
bound for $C_1R_e$; however, in writing Eq.\ (\ref{final1}), we have 
opted to ignore this exceedingly minor detail. Equation (\ref{final1}) 
gives a clear linear dependence of $N$ on $R_e$. Thus, we have 
essentially recovered the bound (\ref{N1}), up to the constant factor 
$C_1$ and a slight difference in the definition of $R_e$, which was 
obtained earlier by counting the active modes from the smallest wave 
number $k_0=1/L$ to the dissipation wave number 
$k_d=\norm{\nabla\omega}/\norm{\omega}$.

\subsection{Degrees of freedom in enstrophy space}

An upper bound for $N$ in the enstrophy space is derived in a similar
manner. From  Eq.\ (\ref{sum2}) we have
\begin{eqnarray}
\label{SUM2}
\sum_{i=1}^{n}\Lambda_i &=&
-\sum_{i=1}^{n}\left(\langle\Delta\vartheta^i J(\vartheta^i,\omega)\rangle 
+ \nu\norm{\nabla\Delta\vartheta^i}^2\right) \nonumber\\ 
&\le&
\sum_{i=1}^{n}\left(\langle|\Delta\vartheta^i||\nabla\vartheta^i|
|\nabla\omega|\rangle - \nu\norm{\nabla\Delta\vartheta^i}^2\right) 
\nonumber\\ 
&\le&
\left\langle\left(\sum_{i=1}^{n}|\Delta\vartheta^i|^2\sum_{i=1}^{n}
|\nabla\vartheta^i|^2\right)^{1/2}|\nabla\omega|\right\rangle
- \nu\sum_{i=1}^{n}\norm{\nabla\Delta\vartheta^i}^2 \nonumber\\
&\le&
\left\langle\left(\sum_{i=1}^{n}|\Delta\vartheta^i|^2\sum_{i=1}^{n}
|\nabla\vartheta^i|^2\right)^2\right\rangle^{1/4}
\langle|\nabla\omega|^{4/3}\rangle^{3/4} 
- \nu\sum_{i=1}^{n}\norm{\nabla\Delta\vartheta^i}^2
\nonumber\\ &\le&
\norm{\sum_{i=1}^{n}|\nabla\vartheta^i|^2}_\infty^{1/2}
\norm{\sum_{i=1}^{n}|\Delta\vartheta^i|^2}^{1/2}
(2\pi L)^{1/2}\norm{\nabla\omega}
- \nu\sum_{i=1}^{n}\norm{\nabla\Delta\vartheta^i}^2, 
\end{eqnarray}
where H\"older's inequalities with the pairs of conjugate exponents $4/3$ 
and $4$ and $3/2$ and $3$ have been used in the penultimate and final 
steps, respectively. For further estimates of the terms in this equation, 
we employ a few more analytic inequalities concerning the orthonormal set 
$\{\vartheta^i\}_{i=1}^n$ in the enstrophy space. First, 
we have \cite{Constantin88,Constantin87}
\begin{eqnarray}
\label{ineq3}
\norm{\sum_{i=1}^{n}|\nabla\vartheta^i|^2}_\infty &\le& 
c_3^2\left(1 + \ln\sum_{i=1}^nL^2\norm{\nabla\Delta\vartheta^i}^2\right),
\end{eqnarray}
where $c_3$ is a non-dimensional constant independent of the set 
$\{\vartheta^i\}_{i=1}^n$. Second, a version of Eq.\ (\ref{ineq1}) for 
the present orthonormal set is
\begin{eqnarray}
\label{ineq4} 
\norm{\sum_{i=1}^{n}|\Delta\vartheta^i|^2} \le 
c_1\left(\sum_{i=1}^n\norm{\nabla\Delta\vartheta^i}^2\right)^{1/2}.
\end{eqnarray}  
Finally, a version of Eq.\ (\ref{ineq2}) for $\{\vartheta^i\}_{i=1}^n$ is 
\begin{eqnarray}
\label{ineq5} 
\sum_{i=1}^n\norm{\nabla\Delta\vartheta^i}^2 \ge \frac{c_2^2}{L^2}n^2.
\end{eqnarray}  

Now by substituting Eqs.\ (\ref{palins}), (\ref{ineq3}), and (\ref{ineq4}) 
into Eq.\ (\ref{SUM2}) we obtain
\begin{eqnarray}
\label{SUM3}
\sum_{i=1}^{n}\Lambda_i &\le&
C'\left(1 + \ln\sum_{i=1}^nL^2\norm{\nabla\Delta\vartheta^i}^2\right)^{1/2}
\left(\sum_{i=1}^nL^2\norm{\nabla\Delta\vartheta^i}^2\right)^{1/4}
\frac{\norm{\omega_0}_\infty^{1/2}\norm{\omega_0}}{\nu^{1/2}}
- \nu\sum_{i=1}^{n}\norm{\nabla\Delta\vartheta^i}^2 \nonumber\\
&=&
\frac{\nu\xi^{1/4}}{L^2}\left(C'\left(1+\ln\xi\right)^{1/2}
\frac{L^2\norm{\omega_0}_\infty^{1/2}\norm{\omega_0}}{\nu^{3/2}}
- \xi^{3/4}\right) =
\frac{\nu\xi^{1/4}}{L^2}\left(C'R_e^{3/2}\left(1+\ln\xi\right)^{1/2}
- \xi^{3/4}\right), \nonumber\\
\end{eqnarray}
where $C'=\sqrt{2\pi}c_1c_3$, 
$\xi=\sum_{i=1}^nL^2\norm{\nabla\Delta\vartheta^i}^2$, and 
$R_e=(L^4\norm{\omega_0}_\infty\norm{\omega_0}^2)^{1/3}/\nu$. Note 
that by Eq.\ (\ref{ineq5}) we have $\xi\ge c_2^2n^2$. Hence without the 
logarithmic term, it would be straightforward to substitute this into 
Eq.\ (\ref{SUM3}) and deduce an upper bound for $N$ similar to Eq.\ 
(\ref{final1}) with the newly defined $R_e$ replacing its previously 
defined (and comparable) counterpart. Since we are interested in the
case $\xi\gg1$, the logarithmic term should introduce a small 
departure to the linear dependence of $N$ on $R_e$ only. In order to 
account for $\ln\xi$, we can ``cover'' it by a fraction of $\xi$, say 
$\xi/2$. By elementary calculus, we find that 
\begin{eqnarray}
C'R_e^{3/2}\left(1+\ln\xi\right)^{1/2} - \frac{\xi^{3/4}}{2}
\le \sqrt{2}C'R_e^{3/2}\left(1+\ln R_e\right)^{1/2},
\end{eqnarray}
where we have dropped a negative term on the right-hand side. It 
follows that
\begin{eqnarray}
\label{noname}
C'R_e^{3/2}\left(1+\ln\xi\right)^{1/2} - \xi^{3/4} &\le& 
\sqrt{2}C'R_e^{3/2}\left(1+\ln R_e\right)^{1/2} - \frac{\xi^{3/4}}{2} 
\nonumber\\
&\le&
\sqrt{2}C'R_e^{3/2}\left(1+\ln R_e\right)^{1/2} - \frac{(c_2n)^{3/2}}{2}.
\end{eqnarray}
The condition $\sum_{i=1}^{n}\Lambda_i \le 0$ is satisfied when the 
right-hand side of Eq.\ (\ref{noname}) is non-positive. This requires a 
straightforward condition for $n$ which in turn yields the result
\begin{eqnarray} 
\label{final2}
N &\le& C_2R_e\left(1 + \ln R_e\right)^{1/3},
\end{eqnarray} 
where $C_2=(8{C'}^2)^{1/3}/c_2$. 

As expected, Eq.\ (\ref{final2}) gives an essentially linear scaling of 
$N$ with $R_e$ since the superlinear dependence on $R_e$, due to the 
logarithmic term, is slight for large $R_e$. Given that the same linear 
scaling was found earlier in the energy space, this is somewhat 
surprising. The reason is that the energy in two-dimensional turbulence 
is predominantly transferred to smaller wave numbers while the enstrophy 
is predominantly transferred to larger wave numbers. This undoubtedly 
implies that the enstrophy dynamics have relatively more degrees of 
freedom than the energy dynamics. Hence, it is somewhat counter-intuitive 
that Eqs.\ (\ref{final1}) and (\ref{final2}) do not differ by much. A
possible explanation is that Eq.\ (\ref{final1}) may not be as optimal 
as Eq.\ (\ref{final2}). Some qualitative support for this possibility 
turns up in the next subsection.

\subsection{Discussion}

In the 1980s, estimates were derived for the Hausdorff dimension 
$D_H$ of the global attractor of the two-dimensional Navier--Stokes 
system driven by a time-independent force $\f$ 
\cite{Babin83,Constantin85,Constantin88}. These estimates have been 
known to be sharp, allowing just minor improvements for 
the attractor dimension in the energy space only \cite{Robinson03,T04}. 
In the present notations, the respective bounds for $D_H$ in the energy 
and enstrophy spaces are given by
\begin{eqnarray} 
\label{d1}
D_H &\le& c'\frac{L\norm{\nabla^{-1}\f}}{\nu^2}
\le c'\frac{L^2\norm{\f}}{\nu^2} = c'G
\end{eqnarray} 
and
\begin{eqnarray} 
\label{d2}
D_H &\le& c''\left(\frac{L^2\norm{\f}}{\nu^2}\right)^{2/3}
\left(1 + \ln \frac{L^2\norm{\f}}{\nu^2}\right)^{1/3} \nonumber\\ 
&=& c''G^{2/3}(1+\ln G)^{1/3},
\end{eqnarray} 
where $c'$ and $c''$ are constant and $G$ is known as the generalised 
Grasshof number. Although $G$ has some certain physical significance, 
its highly superlinear dependence on $\nu^{-1}$ appears to make Eqs.\ 
(\ref{d1}) and (\ref{d2}) in disagreement with the bounds for $N_c$ 
and for $N$ derived earlier. We claim that this 
apparent disagreement is due entirely to the particular form of $\f$ 
and could be fully reconciled. For the remainder of this paper, we 
will elaborate on this claim.  

Due to rigor requirement in the mathematical formulation 
\cite{Babin83,Constantin85,Constantin88}, a time-independent forcing 
$\f$ has been used as a model for energy and enstrophy injection. 
The forced Navier--Stokes equations
\begin{eqnarray}
\label{NS}
\u_t + (\u\cdot\nabla)\u + \nabla p &=& \nu\Delta\u + \f \\
\nabla\cdot\u &=& 0 \nonumber
\end{eqnarray}
then admit the following evolution equations
\begin{eqnarray}
\label{energy}
\frac{1}{2}\frac{d}{dt}\norm{\u}^2 
&=&
-\nu\norm{\nabla\u}^2 + \langle\u\cdot\f\rangle \nonumber\\
&\le&
-\nu\norm{\nabla\u}^2 + \norm{\nabla\u}\norm{\nabla^{-1}\f} \nonumber\\
&\le&
-\frac{\nu}{2}\norm{\nabla\u}^2 + \frac{\norm{\nabla^{-1}\f}^2}{2\nu}
\end{eqnarray}
and
\begin{eqnarray}
\label{enstrophy}
\frac{1}{2}\frac{d}{dt}\norm{\nabla\u}^2 
&=&
-\nu\norm{\Delta\u}^2 - \langle\Delta\u\cdot\f\rangle \nonumber\\
&\le&
-\nu\norm{\Delta\u}^2 + \norm{\Delta\u}\norm{\f} \nonumber\\
&\le&
-\frac{\nu}{2}\norm{\Delta\u}^2 + \frac{\norm{\f}^2}{2\nu}
\end{eqnarray}
for the energy $\norm{\u}^2/2$ and enstrophy $\norm{\nabla\u}^2/2$, 
respectively. In Eqs.\ (\ref{energy}) and (\ref{enstrophy}), the terms
$\norm{\nabla^{-1}\f}^2/(2\nu)$ and $\norm{\f}^2/(2\nu)$ represent 
upper bounds for the energy and enstrophy injection rates, respectively. 
Their dependence on $\nu$ is inescapable because the injection rates 
$\langle\u\cdot\f\rangle$ and $-\langle\Delta\u\cdot\f\rangle$
themselves are flow dependent. 

We now demonstrate how the viscosity dependence of the injection rates
(or more precisely of the upper bounds for the injection rates) contributes 
to the superlinear scaling of $D_H$ with $\nu^{-1}$. To this end, let us 
recall the intermediate steps \cite{Temam97,Constantin85,Constantin88} 
toward (\ref{d1}) and (\ref{d2}) given below
\begin{eqnarray} 
\label{d3}
D_H &\le& c'\frac{L\overline{\norm{\nabla\u}^2}^{1/2}}{\nu}
\end{eqnarray} 
and
\begin{eqnarray} 
\label{d4}
D_H &\le& c''\left(\frac{L^2\overline{
\norm{\Delta\u}^2}^{1/2}}{\nu}\right)^{2/3}
\left(1 + \ln \frac{L^2\overline{\norm{\Delta\u}^2}^{1/2}}
{\nu}\right)^{1/3},
\end{eqnarray} 
where the overline denotes the supremum of an asymptotic average. 
From Eqs.\ (\ref{energy}) and (\ref{enstrophy}) we can deduce the forced 
dissipative balance equations
\begin{eqnarray}
\label{balance1}
\overline{\norm{\nabla\u}^2}^{1/2} \le \frac{\norm{\nabla^{-1}\f}}{\nu}
\end{eqnarray}
and
\begin{eqnarray}
\label{balance2}
\overline{\norm{\Delta\u}^2}^{1/2} \le \frac{\norm{\f}}{\nu}.
\end{eqnarray}
Upon substituting these into Eqs.\ (\ref{d3}) and (\ref{d4}), we recover
Eqs.\ (\ref{d1}) and (\ref{d2}), respectively. 

However, if the driving force could somehow be modelled in such a way 
that the averaged energy and enstrophy injection rates would be bounded 
independently of viscosity, say by $\epsilon^2$ and $\eta^2$, respectively, 
then Eqs.\ (\ref{balance1}) and (\ref{balance2}) would become 
\begin{eqnarray}
\label{balance3}
\overline{\norm{\nabla\u}^2}^{1/2} \le \frac{\epsilon}{\nu^{1/2}}
\end{eqnarray}
and
\begin{eqnarray}
\label{balance4}
\overline{\norm{\Delta\u}^2}^{1/2} \le \frac{\eta}{\nu^{1/2}}.
\end{eqnarray}
Then upon substituting these into Eqs.\ (\ref{d3}) and (\ref{d4}), we 
obtain
\begin{eqnarray} 
\label{d5}
D_H &\le& c'\frac{L\epsilon}{\nu^{3/2}}
\end{eqnarray} 
and
\begin{eqnarray} 
\label{d6}
D_H &\le& c''\left(\frac{L^2\eta}{\nu^{3/2}}\right)^{2/3}
\left(1 + \ln \frac{L^2\eta}{\nu^{3/2}}\right)^{1/3}.
\end{eqnarray} 
One can see that Eq.\ (\ref{d6}) has the desired scaling, i.e., linear
dependence on $\nu^{-1}$ with a logarithmic ``correction'' as in Eq.\
(\ref{final2}). Hence, for the dimension estimate in the enstrophy
space, the ``extra'' dependence on $\nu^{-1}$ would be completely removed 
if the enstrophy injection could be bounded independently of viscosity. 
Note, however, that only part of the extra dependence on $\nu^{-1}$ would 
be removed from the dimension estimate in the energy space when the energy 
injection is made independent of viscosity. This strengthens our earlier 
suggestion that the estimate for $N$ (and $D_H$) in the energy space might 
not be as optimal as its counterpart in the enstrophy space.

Closely related to the present notion of degrees of freedom are the 
concepts of determining modes, nodes, and finite-volume elements 
\cite{Foias83,Foias84,Friz01,Jones92,Jones93}. A large body of research 
on the number of degrees of freedom deduced from these concepts has 
produced upper bounds proportional to $G$ (cf.\ \cite{Jones93}). Like 
the estimate for $D_H$ in the energy space, these bounds would reduce 
to $\propto \nu^{-3/2}$ if the enstrophy injection could be bounded 
independently of viscosity. In another mathematical study \cite{Liu93} 
relevant to the present problem, it has been found that when the 
time-independent $\f$ in (\ref{NS}) consists of a single Fourier mode, 
the unstable manifold emanating from the stationary solution 
$-\Delta^{-1}\f/\nu$ has a dimension not lower than $\propto G^{2/3}$. 
This is a lower bound for the Hausdorff dimension $D_H$ of the global 
attractor, and the superlinear dependence on $\nu^{-1}$ of this bound 
is a consequence of the dependence on $\nu$ of the stationary solution.
  
In passing, it is worth mentioning that numerical simulations of 
two-dimensional turbulence have routinely used a variety of forcing 
that provides steady energy and enstrophy injection rates $\epsilon^2$
and $\eta^2$. This class of forcing includes white noise and flow
dependent forcing \cite{T04',TB04}. While such a class of forcing is 
numerically desirable and realistic in some sense, it may render Eq.\ 
(\ref{NS}) incompatible with the mathematical formulation leading to 
the desired estimate (\ref{d6}). Nevertheless, for the present approach, 
there are no technical difficulties in arriving at this estimate as an
upper bound for the number of degrees of freedom in the present sense.

\section{CONCLUSION}

In conclusion, we have derived upper bounds for the number of degrees 
of freedom $N$ of two-dimensional Navier--Stokes turbulence freely 
evolving from a smooth initial vorticity field in a doubly periodic 
domain. This number is defined as the minimum dimension such that 
arbitrary phase space volume elements of no lower dimensions along
the solution curve in phase space contract exponentially under the 
linearized dynamics. This means that the (locally in time) turbulent 
dynamics could be sufficiently ``contained'' within a linear subspace 
whose dimension does not exceed $N$. In essence, $N$ represents a
reduced dimension that a modelled system should achieve in order to 
describe the turbulence adequately. It is found that $N\le C_1R_e$ in 
the energy space and $N\le C_2R_e(1+\ln R_e)^{1/3}$ in the enstrophy 
space. Here $C_1$ and $C_2$ are constant and $R_e$ is the Reynolds 
number, which is defined in terms of the initial vorticity, the system 
size, and the viscosity. These results are consistent with the number 
of active modes deduced from a recent mathematical estimate of the 
viscous dissipation wave number $k_d=\norm{\nabla\omega}/\norm{\omega}$. 

The present estimates for $N$ have been compared with well-known bounds 
for the Hausdorff dimension $D_H$ of the global attractor in the forced 
case, and the apparent difference between the linear (or nearly so) 
scaling of $N$ with $R_e$ and the highly superlinear dependence of $D_H$ 
on the inverse viscosity $\nu^{-1}$ has been discussed. We have argued 
that the superlinear dependence of $D_H$ on $\nu^{-1}$ is not an intrinsic 
property of the turbulent dynamics and further suggested that this is a 
``removable artifact,'' arising from the use of a time-independent 
forcing as a model for energy and enstrophy injection that drives the 
turbulence. This suggestion has been strengthened by the fact that the 
``extra'' dependence of $D_H$ on $\nu^{-1}$ would be completely removed 
(at least for the estimate of $D_H$ in the enstrophy space) if one could 
model the driving force in such a way that the enstrophy injection rate
does not depend on the viscosity. Such a forcing can be seen to be more
realistic than ones with viscosity dependent input.   

In the present analysis, we simply follow a trajectory starting from 
an arbitrary smooth initial vorticity field in the solution (function) 
space of the two-dimensional Navier--Stokes equations and monitor the 
evolution (under the linearized dynamics) of the volumes of 
$n$-dimensional balls centred on the trajectory. We estimate how large 
$n$ should be to ensure that these volumes contract exponentially. This 
turns out to be equivalent to the method of estimating the Hausdorff 
dimension of the global attractor of the forced system. The present 
approach can be seen to be highly flexible in application. In general, 
it is applicable to either autonomous or non-autonomous, forced or 
unforced, and finite-dimensional or infinite-dimensional systems. There 
are virtually no special requirements, other than existence of solution, 
for the present definition (and method of analysis) of the number of 
degrees of freedom to make sense. In particular, the existence of the
usual Lyapunov exponents is not an issue. Furthermore, there is no need 
for {\it a priori} knowledge of the existence of an attractor (or a 
global attractor), whose generalized dimensions would normally be 
considered as the number of degrees of freedom of the dynamical system 
in question. Given all this, we may apply the present approach to less 
idealized and more realistic dynamical models without risking to 
compromise mathematical rigor.

We thank the staff of the Isaac Newton Institute for Mathematical 
Sciences for their hospitality and support during the program on 
``The Nuture of High Reynolds Number Turbulence,'' when this paper 
was completed.

\end{document}